\documentstyle[preprint,aps,epsf]{revtex}

\newcommand{\be}{\begin{eqnarray}}
\newcommand{\ee}{\end{eqnarray}}
\newcommand{\vp}{\varphi}
\newcommand{\dJ}{\delta J}
\newcommand{\da}{\delta\alpha}
\newcommand{\dad}{\delta\dot{\alpha}}
\newcommand{\Jp}{\omega_{J}}
\newcommand{\dI}{\delta I}

\begin{document}
\draft
\title {Quantum noise in current biased Josephson junction}
\author {Y. Levinson}
\address {Department of
Condensed Matter Physics,\\ The Weizmann Institute of Science,
Rehovot 76100, Israel}

\date {}
\maketitle
\begin {abstract}
Quantum fluctuations in a current biased Josephson junction, described
in terms of the RCSJ-model, are considered. The fluctuations of the
voltage and phase across the junction are assumed
to be initiated by  equilibrium current fluctuations in the shunting resistor.
This corresponds to low enough temperatures, when fluctuations of the
normal current in the junction itself can be neglected.
We used the quantum Langevin
equation in terms of random variables related to the limit cycle of the
nonlinear Josephson oscillator. This allows to go beyond the perturbation
theory and calculate the widths of the Josephson radiation lines.
\end {abstract}

\section*{Introduction}
\label{sec:Intro}

Fluctuations in an equilibrium system are considered to be "classical" if
the fluctuation frequency is lower than the
the system temperature,
$\hbar\omega\ll T$. For higher frequencies the fluctuations are "quantum"
and contain a contribution called zero-point fluctuations (ZPFs). Due to
the ZPFs there are fluctuations even at zero temperature, when the system
is in its ground state \cite{Ga00}.

The intriguing question: can these zero-point fluctuations be detected,
has recently attracted much attention. The answer depends on what is the
detector and what means detection.
 Measuring Casimir forces between two bodies, both at $T=0$,
 is definitely a way to detect ZPFs of the electromagnetic
field surrounding the bodies. On the
other hand  there is no radiation from a body at $T=0$ and an antenna
placed nearby will receive no signal, which means that current ZPFs in the
body  do not radiate.

The situation is even more complicated if the system is not in
equilibrium. Probably the simplest example is a gas of excited atoms in a
cavity with no radiation in it (the temperature of the radiation is zero,
while the temperature of the gas is not). When de-excited, atoms emit
spontaneously cavity photons, which is in fact interaction with ZPFs of the
cavity field. The de-excitation of the atom can be considered as a signal
of the presence of the ZPFs in the cavity.

This paper is stimulated by experiments with nonequilibrium
pumped Josephson junction systems
\cite{Koch81,Yu89,Mo90}, where ZPFs play an important role
at low temperatures.
In  \cite{Koch81}  ZPFs were detected in a dc-current biased
Josephson junction, measuring the low frequency noise
of the voltage $V(t)$ across the junction.
The voltage noise was generated by intrinsic current fluctuations $\dI(t)$
in a shunting resistor $R$. The frequency of the measured voltage
noise (of the order 100 kHz)
is much below the temperature $T$ of the resistor (1.6 K and 4.2 K)
but quantum phenomena are important
 because this low frequency is mixed due to nonlinear effects with the
Josephson frequency  $\Jp$ (between 10 and 500 GHz),
which is comparable or even higher than the resistor temperature.

The theory of quantum fluctuations in a dc-current biased Josephson junction
 was given in \cite{Koch80}, based on the Langevin equation for
the phase difference across the junction $\vp (t)$, where the random
forces are intrinsic current fluctuations $\dI (t)$
 in the resistor. We
decided to revisit the problem because of following reasons.
 In \cite{Koch80} the authors used the theory of classical fluctuations
developed in \cite{Likh72}, simply replacing in the {\it classical} Langevin
equation the  classical spectral density of the random forces,
$(\dI ^2)_{\omega}=T/\pi R$, by its
quantum equivalent, $(\hbar\omega/2\pi R)\coth(\hbar\omega/2T)$. This
quantum spectral density, containing ZPFs,  is generated by the {\it
symmetrized} correlator
$\langle (1/2)[\dI (t')\dI (t)+\dI (t)\dI (t')]\rangle$,
which is, however, not relevant in the problem considered.
This is because using
 the symmetrized correlator for the random forces in the Langevin
equation imply that all the correlators calculated from this equation
are also symmetrized. On the other hand, the symmetrized correlator
of the voltage $V(t)$ do not represent the measured voltage noise.

This problem was addressed
in \cite{Les97,Gav00,Been01}.
 It was shown that if the device, measuring current noise,
 is a "resonator" at zero temperature,
with resonant frequency $\omega$, the measured signal
is proportional to the Fourier component $S(\omega)$
of the {\it nonsymmetrized} current correlator  $\langle j(0)j(t)\rangle$,
with $\omega >0$ and the  dc component in the current
$j(t)$ being subtracted.
$S(\omega)$ is real, but not symmetric in $\omega$,
and hence the condition $\omega>0$ is essential.
If the current
carrying system is in equilibrium at zero temperature, one finds
$S(\omega)=0$ for $\omega>0$, which means there is no signal created by ZPFs.
It is because the signal $S(\omega)$
 is in fact proportional to the power spectral density
{\it spontaneously} radiated by the system.
The situation is similar in quantum optics \cite{Se93}, where ZPFs
can create radiation only in nonequilibrium systems, like parametric
amplifiers.

The second reason to revisit the theory is as follows. In \cite{Likh72}
and \cite{Koch80} the fluctuations were calculated using perturbation
theory for the phase $\vp(t)$ ,
i.e. assuming that $(\delta\vp ^2)_{\omega}$ is small
and proportional
to $(\dI ^2)_{\omega}$. This theory diverges near $\omega=0$
and the harmonics of Josephson frequency
$\omega=k\Jp$, $k=1,2,..$, which are exactly the points of interest.

To account for all the above mentioned circumstances we use the {\it quantum}
Langevin equation for $\vp (t)$, formulating it according to the
general recipe given in \cite{Fo88}.
With no random forces the solution of the Langevin equation
for the phase, $\overline{\vp}(t)=f(\Jp t+\alpha_{0})$,
contains an arbitrary initial phase $\alpha_{0}$, since the equation
for $\overline{\vp}(t)$ is invariant with respect to time shift.
Due to random forces this initial phase acquires a fluctuating
contribution $\da (t)$, and the  Langevin equation can be rewritten
for $\da (t)$. Using this approach we
 avoid the perturbation theory for $\vp(t)$,
 which allows to calculate the shape of the radiation lines
at the Josephson frequency harmonics $k\Jp$.

The width of the Josephson emission lines was considered in the
well known papers \cite{St6869}, however a white spectrum was assumed
for the random forces. This assumption is employed in almost all
papers dealing with noise influence on Shapiro steps \cite{BJ84},
on the impedance of microwave driven junctions \cite{Co00,Fi02}
and other phenomena in Josephson junctions
(more references are given in \cite{Ba82,Bl01}).
There are only few exceptions. In \cite{Be97,Mi96} the white noise
 with correlator $\sim\delta(t)$
was replaced by a dichotomous telegraph noise with
an exponential correlator $\sim\exp(-\gamma t)$.
In \cite{Ge01} the sum of these two was considered.
Full consideration for the quantum correlations, considering the
rounding of the $I-V$- curve, was given in \cite{Me86}.

\section{Quantum Langevin equation}
\label{QLe}
The equations describing a current biased Josephson junction in
the RCSJ-model are as follows \cite{Ti96,Ba82}
\be
\label{meq}
I_{c}\sin\vp(t)+\frac{V(t)}{R}+C\frac{d V(t)}{dt}
=I_{p}+\delta I(t),\qquad
\frac{d\vp(t)}{dt}=\frac{2e}{\hbar}V(t).
\ee
Here $I_{c}$ and $I_{p}$ are the critical and bias currents,
$R$ and $C$ are the resistor and capacitor of the model,
$\vp(t)$ and $V(t)$ are the phase and voltage difference across
the  junction and $\delta I(t)$ are the intrinsic current
fluctuations.
The phase and voltage differences contain
fluctuations generated by $\delta I(t)$ and in
what follows we write $\vp(t)=\overline{\vp}(t)+\delta\vp(t)$
and $V(t)=\overline{V}(t)+\delta V(t)$, separating explicitly the
fluctuating contributions.

We neglect  the quasiparticle current in the junction,
assuming low enough temperatures, which means in terms of the model
that $R$ is just the external shunting resistor.
With this assumption $\delta I(t)$ are the
{\it equilibrium} current fluctuations in the resistor $R$,
the only intrinsic fluctuation source in the RCSJ-model
\cite{Likh72,Ko96}. This simplifies the problem enormously,
since the current fluctuations in the junction are state dependent
(they depend on $\overline{\vp}(t)$) and hence are
 non-stationary for time-dependent voltage $\overline{V}(t)$
\cite{Ko96,Sc85,Ec84}.

We eliminate from
Eqs.(\ref{meq}) the potential $V$, multiply it by $\hbar/2e$ and introduce
$E_{c}=(\hbar/2e)I_{c}$. As a result we have
\be
\label{eqphi}
C\left(\frac{\hbar}{2e}\right)^2\frac{d^2\vp}{dt^2}
+\frac{1}{R}\left(\frac{\hbar}{2e}\right)^2\frac{d\vp}{dt}+
E_{c}\sin\vp-\frac{\hbar}{2e}I_{p}
= \frac{\hbar}{2e}\delta I.
\ee
This equation is a standard form of a Langevin equation for
the phase $\vp$ as the particle "coordinate"  and a "potential"
having  the dimension of energy, the current fluctuations being the random
forces. Comparing with the quantum Langevin equation
in \cite{Fo88} we find the correlator
\be
\label{dIcorr}
\langle\delta I(t')\delta I(t)\rangle=\frac{1}{\pi R}
\int_{0}^{\infty}d\omega\hbar\omega\left[N(\omega)e^{-i\omega(t-t')}
+(N(\omega)+1)e^{i\omega(t-t')}\right]
\ee
and the commutator
\be
\label{dIcomm}
[\delta I(t'),\delta I(t)]=\frac{2i}{\pi R}
\int_{0}^{\infty}d\omega\hbar\omega\sin\omega(t-t')=
-\frac{2i\hbar}{R}\frac{d}{dt}\delta(t-t'),
\ee
where
%\be
%\label{Pl}
$N(\omega)=\left[\exp(\hbar\omega/T)-1\right]^{-1}.$
%\ee
It is important that $\delta I(t)$ is a Gaussian process (in the quantum case
one has to preserve the order of operators) and the commutator Eq.(\ref{dIcomm})
is a $c$-number. (When the thermal bath responsible for the random forces
is chosen to be a collection of harmonic oscillators \cite{Fo88}
the Gaussian properties of the random forces follow from the Gaussian properties
of a harmonic oscillator in equilibrium. However, probably, this is a more general
property, since many degrees of freedom of the thermal bath contribute to
the random force.)

We will use the following notations for Fourier components and spectra
of correlators
\be
A(t)=\int_{-\infty}^{+\infty}d\omega e^{-i\omega t}A _{\omega},
\qquad  (A^{\dag})_{\omega}= (A_{\omega})^{\dag}\equiv A^{\dag}_{\omega},
\ee
\be
\langle A(t')^{\dag} A(t)\rangle=\int_{-\infty}^{+\infty}d\omega e^{-i\omega (t-t')}
(A ^2)_{\omega},
\qquad \langle A_{\omega '}^{\dag}A_{\omega}\rangle=
\delta(\omega-\omega')(A ^2)_{\omega},
\ee
where $^{\dag}$ means "hermitian conjugate".
In these notations the spectrum of equilibrium current fluctuations is
\be
\label{dI}
(\dI ^2)_{\omega}=\frac{\hbar\omega}{\pi R} N(\omega).
\ee
The width of the spectrum is $T/\hbar$ and the corresponding correlation
time is $\hbar/T$.
At $T=0$ this spectrum has no Fourier components with $\omega>0$,
which means that energy can not be extracted from the thermal bath.
The spectral density of the {\it symmetrized} correlator, which contains
ZPFs,
\be
\label{dIsym}
\frac{1}{2}\left[(\dI ^2)_{\omega}+(\dI ^2)_{-\omega}\right]
=\frac{\hbar\omega}{\pi R} \left(N(\omega)+\frac{1}{2}\right)=
\frac{\hbar\omega}{2\pi R}\coth\left(\frac{\hbar\omega}{2T}\right)
\ee
does  not have this property.
In the classical case, when $T\gg\hbar\omega$ and $|t-t'|\gg \hbar/T$,
the random forces have a "white" spectrum, i.e.
\be
\label{dIcl}
(\dI ^2)_{\omega}=\frac{T}{\pi R}\, ,\qquad \langle\dI(t')\dI(t)\rangle=
\frac{2T}{R}\delta(t-t').
\ee

\section{General considerations}
\label{gen}
We recall first the properties of the junction when one neglects
fluctuations. The solution of Eqs.(\ref{meq})
with $\delta I=0$ differs qualitatively
in cases when the bias current $I_{p}$ is smaller or greater than the
critical current $I_{c}$. We will consider only the case of the
non-stationary Josephson effect, when $I_{p}>I_{c}$. In this case
the voltage $\overline{V}(t)$ and the phase $\overline{\vp}(t)$
are  periodic functions of time (the last one modulo $2\pi$),
the period defining the Josephson frequency $\Jp$.
These functions can be presented as follows
\be
\label{lc}
\overline{\vp}(t)=f(x),\qquad
\overline{V}(t)=I_{c}R\,\, g(x),\qquad x=\Jp t+\alpha_{0},
\ee
where $f(x)$ and $g(x)$ are non-dimensional functions
\be
\label{FG}
g(x)=\sum_{k}g_{k}e^{-ikx},\qquad f(x)=\sum_{k}f_{k}e^{-ikx}+x,
\ee
$\alpha_{0}$ is an arbitrary
initial phase, appearing because  Eqs.(\ref{meq}) are invariant with
respect to time shift. Eqs.(\ref{lc}) define the limit cycle of
the nonlinear Josephson oscillator, i.e. the trajectory of the system
in the phase space $(\vp,V)$. The time-average voltage
across the junction is $V_{0}=I_{c}Rg_{0}$.

As was explained in the Introduction the measured voltage noise
(proportional to the radiated energy) is
\be
\label{S}
S(\omega)={1\over 2\pi}\int_{-\infty}^{+\infty}dt e^{i\omega t}
\langle v(0)v(t)\rangle
\ee
with $v(t)=V(t)-V_{0}$ and $\omega>0$.
When fluctuations are neglected, $v(t)$ is a non-random periodic function.
Considering  the initial phase $\alpha_{0}$ to be random
within the interval $(0,2\pi)$, we
convert  the periodic function $v(t)$ into its random equivalent, i.e. a
stationary random process with an equidistant discrete spectrum.
 Using
\be
\label{avalpha}
\overline{\exp{[-i(k-k')\alpha_{0}]}}=\delta_{k,k'},
\ee
we average $v(t')v(t)$ over $\alpha_{0}$ and obtain the relevant correlator
\be
\label{corrmont}
\langle v(t')v(t)\rangle = (I_{c}R)^2
\sum_{k\neq 0}|g_{k}|^{2}\exp[-ik\omega_{J}(t-t')]
\ee
and the noise spectrum
\be
\label{corrmon}
(v^2)_{\omega}\equiv
S(\omega)=(I_{c}R)^2\sum_{k\neq 0}|g_{k}|^{2}
\delta(\omega-k\Jp),\qquad \omega>0.
\ee

We turn now to the effect of fluctuations, when the random forces $\dI(t)$
tend to
move the system away from the limit cycle and destroy the periodicity of
the system dynamics. We assume the random forces  to be weak.
If the system is shifted by the random force
perpendicular to the cycle, a "restoring force" exist, which tends to  bring
the system back to the cycle. However no such force exist when the system
is shifted along the cycle, since this corresponds to a change of
{\it arbitrary} initial phase $\alpha_{0}$.
To account for these two effects we look for a fluctuating solution
of Eqs.(\ref{meq}) in the form
\be
\label{solfl}
\vp(t)=f(\Jp t+\alpha_{0}+\da (t)),
\qquad  V(t)=I_{c}R\;[g(\Jp t+\alpha_{0}+\da (t))+\delta\xi(t)],
\ee
with two unknown random functions  $\da (t)$ and $\delta\xi(t)$
(which are operators in the quantum case). The last one
describes the shift of the system perpendicular to the limit cycle and since
in this case
a restoring force acts against the random forces, $\delta\xi(t)$ is small and
proportional to $\dI (t)$; it contains the same frequencies as $\dI (t)$.
Contrary, $\da (t)$ can grow unlimitedly even for
small $\dI (t)$, if the random force contains the resonance frequency $\Jp$
or its harmonics. Hence, $\da(t)=\overline{\da}(t)+\widetilde{\da}(t)$,
where $\widetilde{\da}(t)$ is proportional to $\dI(t)$
 and contains the frequencies of the random
force, while $\overline{\da}(t)$ is secular and slow,
i.e. contains only low
frequencies proportional to the {\it amplitude} of the random force.
 The function ${\da} (t)$ describes the so called
 diffusion of the initial phase and its
derivative describes Josephson frequency fluctuations,
which are responsible for the broadening of the Josephson radiation lines,
\be
\label{Jpfl}
\delta\Jp(t)=\frac{d}{dt}\da(t)
\equiv\dad(t),\qquad
\da (t)=\int^{t}_{-\infty}dt'\dad (t').
\ee

In terms of the new variables the voltage noise is
\be
\label{vv}
\langle v(t')v(t)\rangle=(I_{c}R)^2\times
\ee
\be
\nonumber
\left\langle
\left[\sum_{k'\neq 0}g_{k'}^{*}\exp[ik'(\Jp t'+\alpha_{0}+\da(t'))]+\delta\xi(t')\right]
\left[\sum_{k\neq 0}g_{k}\exp[-ik(\Jp t+\alpha_{0}+\da(t))]+\delta\xi(t)\right]
\right\rangle.
\ee
%\be
%e^{\displaystyle ik'(\Jp t'+\alpha_{0}+\da(t'))}
%\ee
As we will see later $\{\da(t),\delta\xi(t)\}$ is a Gaussian process and to find the
voltage noise one has to calculate the correlators $\langle\da(t')\da(t)\rangle$,
$\langle\da(t')\delta\xi(t)\rangle$ and $\langle\delta\xi(t')\delta\xi(t)\rangle$.

The effect of the random forces on the noise spectra Eq.(\ref{S})
is twofold: the monochromatic Josephson lines in
Eq.(\ref{corrmon}) acquire a width, and a background for all
frequencies  $\omega>0$ appears. Weak random forces means that
the width of the Josephson lines is small compared to the distance
between them and that the background contribute to the integrated noise
less than the lines do.

Of special interest is the
background  $S(0)$ at low frequencies $\omega\ll\Jp$. To calculate
the background noise at frequencies far enough from the resonant
ones $\omega=k\Jp, k=1,2,...$ one can employ perturbation theory
assuming $\da (t)$ to be small, while to find the width of these
lines a more sophisticated approach has to be used.

\section{Langevin equations for an overdamped junction}

Substituting Eqs.(\ref{lc}) into  Eqs.(\ref{meq}) with $\dI =0$ we find
the relations between the functions $f$ and $g$ as follows
\be
\label{gf}
g(x)+\sin f(x)+\Jp\tau g'(x)=p,\qquad \Jp f'(x)=\omega_{0}g(x),
\ee
where $p=I_{p}/I_{c}$, $\omega_{0}=(2e/\hbar)I_{c}R$,  and $\tau=RC$.
To obtain the Langevin equations for $\delta\xi(t)$ and $\da(t)$
we substitute
 Eqs.(\ref{solfl}) into Eqs.(\ref{meq}) and use Eqs.(\ref{gf}).
As a result we find \be \label{Le}
\tau\delta\dot{\xi}(t)+\delta\xi(t)+\tau g'(x)\dad(t)=\dI
(t)/I_{c}, \qquad f'(x)\dad(t)=\omega_{0}\delta\xi(t) \ee with
$x=\Jp t+\alpha_{0}+\da(t)$. (Some caution is needed in the
quantum case since $\da$ and $\dad$ do not commute. However the
main (secular) part of $\da$ is a slow function of $t$ and
hence $\da(t)$ and $\dad (t')$ effectively commute when $t'=t$.)

In what follows we will consider the overdamped Josephson junction,
assuming the capacitance $C$ to be small, i.e. $\omega_{0}\tau\ll 1$.
 In this case we can skip
in the first of Eqs.(\ref{Le}) the term with $\dad(t)$, since, according
to the second of these equations, it is a small correction to the term
$\delta\xi(t)$. We can also use the functions $f(x)$ and  $g(x)$ as for a
junction with $C=0$, in which case \cite{Ko96} the Josephson frequency
 is $\Jp=\omega_{0}\sigma$ with $\sigma=(p^{2}-1)^{1/2}$ and
\be
\label{g}
g(x)=(p^{2}-1)/(p+\sin x);
\qquad
g_{k}=(-i)^{k}\sigma(p-\sigma)^{k},\quad k\geq 0;\qquad g_{-k}=g_{k}^{*}.
\ee
The functions $f(x)$ and $g(x)$ for an overdamped junction are related as follows
\be
\label{fg}
\sigma f'(x)=g(x), \qquad g(x)+\sin f(x)-p=0.
\ee
As a result we have the following Langevin equations
\be
\label{daxi}
\qquad \tau\delta\dot{\xi}(t)+\delta\xi(t)=\frac{\dI (t)}{I_{c}},
\qquad
\dad (t)=\frac{\omega_{0}}{\sigma}[p+\sin(\Jp t+\alpha_{0}+\da (t))]\;\delta\xi(t).
\ee
From the first of these equations we find immediately
\be
\delta\xi_{\omega}=\frac{1}{I_{c}}\frac{\dI_{\omega}}{1-i\omega\tau}
\equiv\frac{\dJ_{\omega}}{I_{c}},
\qquad (\dJ ^2)_{\omega}=\frac{(\dI ^2)_{\omega}}{1+(\omega\tau)^2},
\ee
where $\dJ$ are current fluctuations shunted by the capacitor.
As mentioned already
$\da(t)=\overline{\da}(t)+\widetilde{\da}(t)$, where $\widetilde{\da}(t)$
is fast oscillating but small, while $\overline{\da}(t)$ is slow but
generally not small.
When performing the Fourier transform of $\dad(t)$
in the second of Eqs.(\ref{daxi}) one can neglect
in the right hand side the small
$\widetilde{\da}(t)$ and replace the slow $\overline{\da}(t)$ by a constant,
absorbed in $\alpha_{0}$.
With this approximation
\be
\label{dadf}
\dad_{\omega}=\frac{\omega_{0}}{I_{c}}
\left\{\frac{p}{\sigma}\dJ_{\omega}+\frac{1}{2i\sigma}
\Big[\exp[i\alpha_{0}]\dJ_{\omega+\Jp}-
\exp[-i\alpha_{0}]\dJ_{\omega-\Jp}     \Big]      \right\}.
\ee
If one takes into account the slow variation of $\overline{\da}(t)$, then
$\dJ_{\omega}$ is replaced by its convolution with the narrow spectra of
$\overline{\da}(t)$, but this effect can be neglected for weak enough
random forces.
In the time domain Eq.(\ref{dadf}) corresponds to
\be
\label{dadt}
\dad (t)=\frac{\omega_{0}}{\sigma}[p+\sin(\Jp t+\alpha_{0})]
\frac{\dJ (t)}{I_{c}}.
\ee
To calculate $(\dad^2)_{\omega}$ we average
$\langle\dad_{\omega '}^{\dag}\dad_{\omega}\rangle$
over $\alpha_{0}$ using Eq.(\ref{avalpha})
and obtain
\be
\label{dad2}
(\dad^2)_{\omega}=\left(\frac{\omega_{0}}{I_{c}}\right)^{2}
\left\{\left(\frac{p}{\sigma}\right)^{2}(\dJ^2)_{\omega}+
\frac{1}{4\sigma^2}\left[(\dJ^2)_{\omega+\Jp}+(\dJ^2)_{\omega-\Jp}\right]
\right\}.
\ee
It is convenient to present
\be
\label{1overPhi}
\frac{1}{f'(x)}=\sum_{k=-\infty}^{\infty}b_{k}e^{-ikx},
\qquad b_{0}=\frac{p}{\sigma}; \qquad b_{\pm 1}=\pm\frac{1}{2i\sigma};
\qquad b_{k}=0,\;\; (|k|>1).
\ee
Using this representation and
the intrinsic current fluctuation spectrum given by Eq.(\ref{dI})
we have
\be
\label{dad}
(\dad^2)_{\omega}=\frac{4}{\pi g}\sum_{k}|b_{k}|^2
\left\{\frac{\omega N(\omega)}{1+(\omega\tau)^2}\right\}
_{\displaystyle\omega\Rightarrow\omega-k\Jp},
\ee
where $g=(\hbar/e^2)R^{-1}$ is the non-dimensional conductance of the
junction resistor.
It what follows we assume $g\gg 1$, and we will see that this is the main
condition for the fluctuations to be weak.
We will not consider very strong
pumping and assume $p\simeq 1$. Then it follows from
$\omega_{0}\tau\ll 1$  that also $\Jp\tau\ll 1$
and one can simplify Eq.(\ref{dad}),
\be
\label{dadm}
(\dad^2)_{\omega}=\frac{4}{\pi g}\Pi(\omega)\sum_{k}|b_{k}|^2
\left\{\omega N(\omega)\right\}
_{\displaystyle\omega\Rightarrow\omega-k\Jp},
\qquad
\Pi(\omega)=\frac{1}{1+(\omega\tau)^2},
\ee
where $\Pi(\omega)$ is the shunting factor. It is obvious from Eq.(\ref{dad2})
that approaching the threshold $p=1$ of the non-stationary Josephson effect, when
$\sigma\rightarrow 0 $, the fluctuations increase.

\section{Perturbation theory}
We consider shortly the situation when $\da(t)$ can be assumed to be small.
As we will see no problems appear in perturbation theory
if one neglects the capacitance of the junction and put $\tau=0$.
Substituting in the second of Eqs.(\ref{solfl}) the Fourier expansion of $g$
from Eqs.(\ref{FG}) and
expending in $\da(t)$ we find the voltage fluctuations
\be
\label{dVt}
\delta V(t)= V(t)-\overline{V}(t)
=I_{c}R\sum_{k}(-ik)g_{k}\exp[-ik(\Jp t+\alpha_{0})]\da(t)+
R\dI(t).
\ee
Using $\da_{\omega}=\dad_{\omega}/(-i\omega)$ we have for
its Fourier transform
\be
\label{dVomega}
\delta V_{\omega}=I_{c}R\sum_{k}kg_{k}\exp[ -ik\alpha_{0}]
\frac{\dad_{\omega-k\Jp}}{\omega-k\Jp}+R\dI_{\omega}.
\ee
We substitute here Eq.(\ref{dadf}) and shift the summation over $k$, obtaining
\be
\delta V_{\omega}=\sum_{k}\exp[-ik\alpha_{0}]Z_{k}(\omega)\dI_{\omega-k\Jp},
\ee
where the impedance of the junction is \cite{Likh72}
\be
Z_{k}(\omega)=R\left\{\delta_{k,0}+\frac{p}{\sigma}\frac{kg_{k}}{\omega-k\Jp}+
\frac{1}{2i\sigma}\left[\frac{(k+1)g_{k+1}}{\omega-(k+1)\Jp}-
\frac{(k-1)g_{k-1}}{\omega-(k-1)\Jp}\right]\right\}.
\ee
Averaging over the initial phase $\alpha_{0}$
we have the general result of the perturbation
theory as follows
\be
\label{pt}
(\delta V^2)_{\omega}=\sum_{k}|Z_{k}(\omega)|^2 (\dI ^2)_{\omega-k\Jp},
\qquad \omega>0.
\ee
 In the classical case  one can see, using Eq.(\ref{dIcl}), that
the above result agrees with \cite{Likh72}.
However in the quantum case
the voltage noise is not given by the symmetrized current
spectral density, as it was obtained in \cite{Koch80}.
It follows from Eq.(\ref{dVomega}) that perturbation theory diverges
at $\omega=k\Jp$ for $k=1,2,...$, i.e. at the frequencies of the Josephson
lines. This is because, as already mentioned, at these frequencies $\da(t)$
is not small. Note, that in \cite{Likh72}, where perturbation theory was
developed for $\delta\varphi(t)$, there was also an artifact divergency for
$(\delta\varphi^2)_{\omega}$ at $\omega=0$.

Using $g_{k}$ from
Eq.(\ref{g}) one can check that $Z_{k}(0)=0$ when $|k|>1$ and find the
low frequency $\omega\ll \Jp$ voltage noise to be
\be
\label{S0}
S(0)=
(\delta V^2)_{0}=|Z_{0}(0)|^2 (\dI^2)_{0}+|Z_{1}(0)|^2[(\dI^2)_{+\Jp}+(\dI^2)_{-\Jp}]
\ee
with
$
|Z_{0}(0)|^2=R^2(p/\sigma)^2,\;\; |Z_{1}(0)|^2=R^2/4\sigma^2,
$
giving
\be
S(0)=\frac{RT}{\pi}\frac{p^2}{\sigma^2}\left[1+\frac{1}{2p^2}\frac{\hbar\Jp}{2T}
\coth\left(\frac{\hbar\Jp}{2T}\right)\right].
\ee
In this special case the symmetrized current spectral noise enters, and
this is why our result for $S(0)$ agrees with that obtained in \cite{Koch80}.

\section{Dephasing factor of Josephson lines}

To find the width of the Josephson lines one has to go beyond the perturbation
theory to eliminate the singularities of $(\delta V^2)_{\omega}$ near
the Josephson frequency and its harmonics. If $\delta V_{\omega}$
is calculated from Eq.(\ref{dVomega}) in the vicinity of $k\Jp, (k=1,2...$),
the term $R\dI_{\omega}$ can be neglected, and
$(\delta V^2)_{\omega}$ is a double sum over $k$ and $k'$.
The strongest singularities
$(\omega-k\Jp)^{-2}$ emerge from the diagonal terms with $k'=k$,
while the non-diagonal terms create weaker singularities $(\omega-k\Jp)^{-1}$.
Comparing the diagonal and non-diagonal terms reveal that when
$|\omega-k\Jp|\ll \Jp$ the diagonal terms dominate.
Hence, to find the width of the Josephson lines $k\Jp$
on can neglect in Eq.(\ref{vv}) the terms $\delta\xi(t)$ and $\delta\xi(t')$
and pick-up in
the double sum over $k$ and $k'$ the terms with $k'=k$ only, considering
\be
\langle v(t')v(t)\rangle_{k} = (I_{c}R)^2 |g_{k}|^2
\exp[-ik\Jp(t-t')]\left\langle \exp[ik\da(t')]\exp[-ik\da(t)]
\right\rangle.
\ee
As a result the delta-function in the spectrum Eq.(\ref{corrmon})
is replaced by a form-factor
\be
\label{ff}
\Phi_{k}(\nu)=
\int_{0}^{\infty}\frac{dt}{2\pi}\exp[i\nu t]D_{k}(t)+c.c.,
\ee
where $\nu=\omega-k\Jp$ is the detuning from the center of the line
and the dephasing factor is
\be
\label{D}
D_{k}(t-t')=\langle\exp[ik\da (t')]\exp[-ik\da (t)]\rangle=D_{k}(t'-t)^*.
\ee
The symmetry relation for the form-factor
follows from the invariance of the average with respect to time shift.
Since, as assumed, the fluctuations $\delta I(t)$ are
small, the dephasing factors decay  slowly with $t$.

As follows from Eq.(\ref{dadt}) and the
properties of $\dI (t)$ described in sec.\ref{QLe}, that  $\dad(t)$ and $\da(t)$
are Gaussian processes and their
commutators are $c$-numbers. Using the well known identity
$
 e^{A}a^{B}=e^{A+B}e^{\frac{1}{2}[A,B]}
$
one can write the operator product entering the dephasing factor Eq.(\ref{D}) as
\be
\exp[ik\da (0)]\exp[ -ik\da (t)]=
\exp[ik(\da(0)-\da(t))]
\exp\left[\frac{1}{2}k^2\,[\da(0),\da(t)]\right].
\ee
Using the Gaussian properties we calculate the averages and
have for the dephasing factor
\be
D_{k}(t)=\exp\left[k^2\,C(t)\right]\exp\left[-k^2\,K(t)\right],
\ee
where
\be
\label{CK}
C(t)&=&\frac{1}{2}[\da(0),\da(t)]=
\frac{1}{2}\int_{-\infty}^{0}dt_{1}\int_{-\infty}^{t}dt_{2}
[\dad(t_{1}),\dad(t_{2})],\\ \nonumber
K(t)&=&\frac{1}{2}\langle[\da(0)-\da(t)]^2\rangle=\frac{1}{2}
\left\langle\left[\int_{0}^{t}dt'\dad(t')\right]^2\right\rangle=\frac{1}{2}
\int_{0}^{t}dt_{1}\int_{0}^{t}dt_{2}\langle\dad(t_{1})\dad(t_{2})\rangle.
\ee
Presenting the commutator entering  $C(t)$ as its average
$\langle\dad(t_{1})\dad(t_{2})-\dad(t_{2})\dad(t_{1})\rangle $,
and substituting the Fourier representation of the correlator
$\langle\dad(t_{1})\dad(t_{2})\rangle$, we find
\be
K(t)=\int_{0}^{\infty}d\omega\;g(\omega)
\frac{1-\cos\omega t}{\omega^2},
\qquad
C(t)=i\int_{0}^{\infty}d\omega\;q(\omega)
\frac{\sin\omega t}{\omega^2}
\ee
with
\be
g(\omega)=\left[(\dad^2)_{\omega}+(\dad^2)_{-\omega}\right],
\qquad
q(\omega)=-\left[(\dad^2)_{\omega}-(\dad^2)_{-\omega}\right].
\ee
Using Eq.(\ref{dadm}) and the relation
$(-\omega) N(-\omega)=\omega(N(\omega)+1)$
one can check that
\be
\label{dad-}
(\dad^2)_{-\omega}=\frac{4}{\pi g}\Pi(\omega)
\sum_{k}|b_{k}|^2\left\{\omega [N(\omega)+1]\right\}
_{\displaystyle \omega\Rightarrow \omega-k\Jp},
\ee
and as a result,
\be
\label{dads}
g(\omega)=\frac{4}{\pi g}\Pi(\omega)
\sum_{k}|b_{k}|^2\left\{\omega\,[2N(\omega)+1]\right\}
_{\displaystyle \omega\Rightarrow \omega-k\Jp},
\ee
\be
\label{dadas}
q(\omega)=\frac{4}{\pi g}\Pi(\omega)
\sum_{k}|b_{k}|^2
\left\{\omega\right\}
_{\displaystyle \omega\Rightarrow \omega-k\Jp}=
\frac{4}{\pi g}\frac{2p^2+1}{2\sigma^2}\omega \Pi(\omega).
\ee
 $q(\omega)$ contains only ZPFs and is temperature independent.
 This is not the case for $g(\omega)$, which for
  high and low temperatures is, correspondingly,
(for $\omega>0$)
\be
\label{gomega}
g(\omega)&=&\frac{4}{\pi g}\frac{2p^2+1}{2\sigma^2}\Pi(\omega)\;
\omega\coth\left(\frac{\hbar\omega}{2T}\right),
\\ \nonumber
g(\omega)&=&\frac{4}{\pi g}\frac{1}{2\sigma^2}\Pi(\omega)
[\Theta(\Jp-\omega)(\Jp-\omega)+(2p^2+1)\omega],
\ee
where $\Theta(\omega)$ is the step function.
Because of the ZPFs $g(\omega)\neq 0$ at $T=0$.
The shunting cut-off is not crucial in calculating $C(t)$; this integral
converge even if one replace $\Pi(\omega)\Rightarrow 1$. Contrary, the integral
$K(t)$ converges only due to this cut-off.

\section{Quantum dynamical narrowing}

As can be seen from the previous section the form-factor of the Josephson
 line is given as an integral of the type
\be
\label{ffg}
\Phi(\nu)=
\int_{0}^{\infty}\frac{dt}{2\pi}\exp[i\nu t]D(t)+c.c.
=\int_{0}^{\infty}\frac{dt}{\pi}\cos[\nu t+H(t)]\exp[-W(t)],
\ee
where the dephasing factor is
$
\label{deffg}
D(t)=\exp[-W(t)+i H(t)].
$
The functions $W(t)$ and $H(t)$ are given by their  spectral representations
\be
\qquad W(t)=\int_{0}^{\infty}d\omega G(\omega)\frac{1-\cos\omega t}{\omega^2},
\qquad
H(t)=\int_{0}^{\infty}d\omega Q(\omega)\frac{\sin\omega t}{\omega^2},
\ee
where both spectral densities $G(\omega),\;Q(\omega)$ contain a small pre-factor
(due to $g^{-1}$) and\newline
$ G(\omega)>0$,
while ${\rm sign}\;Q(\omega)={\rm sign}\;\omega,\quad Q(0)=0$.
This is the quantum version of spectral line
dynamical narrowing. (In the classical theory
 $H(t)=0$, hence the line is symmetric in $\nu$,
and $G'(0)=0$. One can see from Eqs.(\ref{gomega}) that this property
indeed holds for high temperatures but is not valid for zero temperature).

The large time asymptotic of $W(t)$ is obtained replacing $G(\omega)$ by $G(0)$,
giving
\be
\label{gammaL}
W(t)=\gamma t, \qquad \gamma=\frac{\pi}{2}G(0).
\ee
Since $Q(0)=0$ the function $H(t)$ is finite at $t\rightarrow\infty$.
If one replace $W(t)$ by its large time asymptotic and neglects $H(t)$,
the form factor becomes a Lorenzian
\be
\label{Lorenz}
\Phi(\nu)=\frac{1}{\pi}\frac{\gamma}{\gamma^2+\nu^2}.
\ee
As we will see, this line shape is valid only for small enough $\nu$.
To get an expression valid also for large $\nu$ and to estimate the
non-Lorenzian contributions we proceed as follows.
We present
\be
W(t)=\gamma t +\tilde{W}(t),\qquad\tilde{W}(t)=
\int_{0}^{\infty} d\omega [G(\omega)-G(0)]\frac{1-\cos\omega t}{\omega^2},
\ee
 and expand
$
\exp\left[-\tilde{W}(t)\right]=1-\tilde{W}(t).
$
Neglecting second and higher order terms in $\tilde{W}(t)$ and $H(t)$,
we  present the form-factor as a sum
of a symmetric and antisymmetric contributions
\be
\Phi(\nu)=\Phi_{s}(\nu)+\Phi_{a}(\nu)
\ee
with
\be
\Phi_{s}(\nu)=\frac{1}{\pi}\frac{\gamma}{\nu^2+\gamma^2}-
\int_{0}^{\infty}\frac{dt}{\pi}\exp[-\gamma t]
\cos\nu t \;\tilde{W}(t),
\\
\Phi_{a}(\nu)=-\int_{0}^{\infty}\frac{dt}{\pi}\exp[-\gamma t]
\sin\nu t \;H(t).
\ee
To calculate the non-Lorenzian parts of the form-factor introduce functions
\be
\Delta_{c}(\omega,\nu)=-\frac{2}{\pi}\int_{0}^{\infty}dt
\cos\nu t \exp[-\gamma t](1-\cos\omega t)=
\\ \nonumber
\frac{2}{\pi}\frac
{\gamma\omega^2(3\nu^2 - \omega^2-\gamma^2 )}
{(\gamma^2 + \nu^2)(\gamma^2 + (\nu - \omega)^2)
 (\gamma^2 + (\nu + \omega)^2)},
\ee
\be
\Delta_{s}(\omega,\nu)=\frac{2}{\pi}\int_{0}^{\infty}dt
\sin\nu t \exp[-\gamma t]\sin\omega t=
%\\ \nonumber
\frac{2}{\pi}\frac
{2\gamma\omega\nu}
{(\gamma^2 + (\nu - \omega)^2)(\gamma^2 + (\nu + \omega)^2)}.
\ee
With this definitions we have
\be
\Phi_{s}(\nu)=\frac{1}{\pi}\frac{\gamma}{\nu^2+\gamma^2}+
\frac{1}{2}\int_{0}^{\infty}\frac{ d\omega}{\omega^2}
[G(\omega)-G(0)]\Delta_{c}(\omega,\nu).
\ee
\be
\Phi_{a}(\nu)=
-\frac{1}{2}\int_{0}^{\infty}\frac{ d\omega}{\omega^2}
Q(\omega)\Delta_{s}(\omega,\nu).
\ee
For small $\gamma$ in the vicinity of $\omega=\nu>0$ both $\Delta_{c,s}(\omega,\nu)$
are smeared $\delta$-functions,
\be
\Delta_{c,s}(\omega,\nu)=\frac{1}{\pi}\frac{\gamma}{(\omega-\nu)^2+\gamma^2}.
\ee
The main assumption of the dynamical narrowing theory
is that due to the small pre-factor
in $G(\omega)$ the Lorenzian width $\gamma$ is small, so that $G(\omega)$ and
$Q(\omega)$ are smooth on the scale $\gamma$. With this assumption one
 can  replace $\Delta_{c,s}(\omega,\nu)$ by $\delta(\omega-\nu)$
for $\nu\gg\gamma$ (because of the factor $\omega^{-2}$). Then for $\nu>0$
one finds
\be
\label{Phi1}
\Phi_{s}(\nu)=\frac{1}{\pi}\frac{\gamma}{\nu^2+\gamma^2}+\frac{G(\nu)-G(0)}{2\nu^2},
\qquad
\Phi_{a}(\nu)=-\frac{Q(\nu)}{2\nu^2}.
\ee
On the other hand,  at $\nu\gg\gamma$
 the Lorenzian and the term with $G(0)$ cancel each other and we find that
 in the far wings of the line the form-factor follows the spectral densities
 entering the dephasing factor, i.e.
\be
\label{wings}
\Phi_{s}(\nu)=\frac{G(\nu)}{2\nu^2}, \qquad
\Phi_{a}(\nu)=-\frac{Q(\nu)}{2\nu^2},
\qquad \nu>0,\;\nu\gg\gamma.
\ee
It is important to note, that this is in fact a "perturbation theory" result
and can be obtained expanding the dephasing factor in
$W(t)$ and $H(t)$. Within the Lorenzian line, when
$\nu\lesssim\gamma$, the estimates for the non-Lorenzian contributions
in Eq.(\ref{Phi1})
can be obtained replacing in the denominators $\nu^2$ by $\gamma^2$.
They are negligible for small enough $\nu$, when
\be
\label{cL}
G(\nu)-G(0),\; Q(\nu)\ll G(0).
\ee

If one can define a scale $\Delta\omega$, which is the spectral width of
$G(\omega)$ and $Q(\omega)$, the results can be summarized in a simple way.
The dynamical narrowing theory is valid when $\gamma\ll\Delta\omega$.
For detuning much smaller than the spectral width,
$\nu\ll\Delta\omega$, the line form-factor is Lorenzian according to
 Eqs.(\ref{Lorenz}) and (\ref{gammaL}).
When the detuning is of the order or larger than the spectral width,
$\nu\gtrsim\Delta\omega$, the form-factor follows the spectral density
according to the perturbation theory result given by
 Eqs.(\ref{wings}). In this
domain the line is asymmetric, the red wing $(\nu<0)$ being enhanced
compared to the blue one $(\nu>0)$. This is because of ZPFs, which enhance
the probability for the oscillator to loss energy  compared to the
probability to gain it, due to the interaction with the thermal bath. This
asymmetry is related to the perturbation theory result Eq.(\ref{pt}),
where the non-symmetrized current fluctuation spectral density enters.

\section{The width of Josephson lines}

Using the results of the previous section we can find the width and the shape
of the Josephson lines. For the line $k\Jp$ one has to put
\be
\label{GQ}
G(\omega)=k^2\;g(\omega),\qquad Q(\omega)=k^2\;q(\omega),
\ee
and as a result the Lorenzian width of this line is according to Eqs.(\ref{gammaL})
\be
\label{gammak}
\gamma_{k}=k^2\Gamma,\qquad \hbar\Gamma=\frac{4}{g\sigma^2}
\left[p^2\;T+\frac{1}{4}\hbar\Jp\coth\left(\frac{\hbar\Jp}{2T}
\right)\right].
\ee

As was mentioned already, our basic assumption about the weakness of the current
fluctuations means that the line width is small compared to the line separation,
which means that the result given by Eqs.(\ref{gammak}) is correct when
\be
\label{mc}
\gamma_{k}\ll\Jp.
\ee
We will show now that this restriction, together with  $p\simeq 1$,
ensures that in the vicinity of the Josephson frequency, i.e. at
$|\omega-k \Jp|\ll \Jp$,
the dynamical narrowing theory is valid and the form-factor of the
Josephson line is a symmetric
Lorenzian with the width given by Eqs.(\ref{gammak}).
The non-Lorenzian contributions and the asymmetry of the line appears
only at $|\omega-k \Jp|\simeq \Jp$, but in this domain one has to employ perturbation
theory including non-diagonal terms with $k'\neq k$.
Two things have to be checked.
First, that the Lorenzian width $\gamma_{k}$ is small compared to the scales of
the spectral densities $G(\omega)$ and $Q(\omega)$, given by Eqs.(\ref{GQ}).
Second, the conditions Eqs.(\ref{cL}), i.e. that
the non-Lorenzian contributions to the form-factor are small.

As can be seen from the results of dynamical narrowing  theory,
the form-factor at given
detuning $\nu$ do not depend on the behavior of the
spectral densities $G(\omega)$ and $Q(\omega)$ at $\omega>\nu$. Since we
are interested only in $|\omega-k \Jp|\ll \Jp$, the screening factor $\Pi(\omega)$ in
$g(\omega)$ and $q(\omega)$ can be neglected.
 As a result, it follows from Eq.(\ref{gomega}) that the scale of
$G(\omega)$ is the larger of $T/\hbar$ and $\Jp$. The situation is more
complicated for $Q(\omega)$ which is linear in $\omega$ and has no scale.
However, Eq.(\ref{Phi1}) for the asymmetric contribution to the form-factor
is still correct for $\nu\gg\gamma$, since the effective scale of a linear
$Q(\omega)$ is $\omega$.

Consider first the situation of high temperatures, $T\gg\hbar\Jp$, when
the scale of $G(\omega)$ is $\Delta\omega=T/\hbar$, and since  $T/\hbar\gg\Jp$,
it follows
from Eq.(\ref{mc}), that this scale is large compared to $\gamma_{k}$.
Using Eq.(\ref{gomega}) and Eq.(\ref{dadas}) one can check
 that the conditions Eq.(\ref{cL}) are satisfied when $\nu\ll T/\hbar$,
 which is valid automatically since $\nu\ll\Jp\ll T/\hbar$.
In  the case of low temperatures, $T\ll\hbar\Jp$,
the scale of $G(\omega)$ is
$\Delta\omega=\Jp$, which is large compared to $\gamma_{k}$
because of Eq.(\ref{mc}).
The conditions Eq.(\ref{cL}) require $\nu\ll \Jp$, which is also satisfied.

The width at high temperatures,
$
\Gamma=(g\sigma^2)^{-1}\; 2(2p^2+1)(T/\hbar),
$
was obtained in \cite{Likh72} considering the singular term $A_{k}/(\omega-k\Jp)^{2}$
in $(\delta V^2)_{\omega}$ calculated from perturbation theory, Eq.(\ref{pt}),
as a wing of a Lorenzian line and assuming that the total
energy radiated in a this line is not influenced by the fluctuations.
In the high temperature case the fluctuations can be considered to be
 weak, Eq.(\ref{mc}), if
$
g\sigma^3\gg k^2 (T/\hbar\omega_{0}).
$

At low temperatures
$
 \Gamma=(g\sigma^2)^{-1}\Jp,
$
and Eq.(\ref{mc}) reduces to
$
g\sigma^2\gg k^2.
$
Finite dephasing of the Josephson lines exist at zero temperature due to
the ZPFs, which are active because the junction is far from equilibrium.

One can see from the above estimates that the fluctuations are weak due to the
large conductance $g$ of the shunting resistor.
Approaching to the threshold $p=1$, when $\sigma\rightarrow 0$, the
fluctuations increase. The effect of the fluctuations is stronger for
high harmonics, $k\gg 1$.

It is important to note
 that the general quantum result for the line width, Eq.(\ref{gammak}),
can not be obtained from the high temperature classical one simply
replacing the temperature $T$  by its
quantum equivalent, $(\hbar\omega/2)\coth(\hbar\omega/2T)$.

We note also that a general relation is valid,
$
\hbar\Gamma=(4\pi e^2/\hbar)S(0)
$,
which is a simple relation between the width of main Josephson line at
$\omega=\Jp$ and the low frequency noise.
As was mentioned in sec.\ref{gen}, there are two indications, pointing
to the smallness of the fluctuations: the ratio of the line width to the
line separation $\Gamma/\Jp$,
and the ratio of the background contribution  to the contribution
of the lines, which can be estimated as $ S(0)\Jp/(g_{0}I_{c}R)^2$.
It follows from the relation between $\Gamma$ and $S(0)$ that
both ratios are of the same order, i.e. that both ways to estimate the
strength of fluctuations are equivalent.

\section{Acknowledgements}

I acknowledge Y. Imry and U. Gavish for many discussions related
to zero-point fluctuations. This work was supported by the Israel Academy
of Sciences.

\end{document}